# An Extended Stochastic Model for Quantitative Security Analysis of Networked Systems

Maochao Xu[1] and Shouhuai Xu[2]
[1] Department of Mathematics, Illinois State University
mxu2@ilstu.edu
[2] Department of Computer Science, University of Texas at San Antonio
shxu@cs.utsa.edu

**Abstract**

Quantitative security analysis of networked computer systems is one of the decades-long open problems in computer security. Recently, a promising approach was proposed in [19], which however made some strong assumptions including the exponential distribution of, and the independence between, the relevant random variables. In this paper, we substantially weaken these assumptions while offering, in addition to the same types of analytical results as in [19], methods for obtaining the desired security quantities in practice. Moreover, we investigate the problem from a higher-level abstraction, which also leads to both analytical results and practical methods for obtaining the desired security quantities. These would represent a significant step toward ultimately solving the problem of quantitative security analysis of networked computer systems.

## 1 Introduction

Quantitative security analysis of networked computer systems from a whole-system, rather than from an individual component, perspective is one of the most important open problems in cyber security. Recently, a promising stochastic process approach to tackling this problem was proposed in [19]. In this approach, the networked computer system is modeled as a complex network/graph $\mathsf{G} = (\mathsf{V}, \mathsf{E})$, where $\mathsf{V}$ is the vertex/node set and $\mathsf{E}$ is the edge set over which direct attacks can be carried out. Then, a stochastic abstraction of the interactions between the attacker and the defender taking place in $\mathsf{G}$ is considered. The main research task is to derive the following two security measures:

- $q(v)$: It is the probability that a node $v \in \mathsf{V}$ is compromised (or attacked) in the steady state.

- $q$: It is the probability that a uniformly-chosen node in $\mathsf{G}$ is compromised (or attacked) in the steady state.

The quantity $q$ is particularly important for decision-making because it captures/reflects the global security level of a networked system in a useful way. For example, given $q$, the defender can design/deploy appropriate cryptographic/security schemes to tolerate the damage of the attacks. (Some extensions to $q$ will be discussed in Section 7.)

**Our contributions.**  In this paper, we make substantial improvements upon [19], which assumed that the relevant random variables follow the exponential distribution and are independent of each other. Specifically, we consider two cases: (I) the case the lower-level attack-defense process is observed, which is the same as in [19] but with much weaker assumptions; (II) the case the higher-level Alternating Renewal process, but not the lower-level attack-defense process, is observed. The latter case was not investigated in [19] and may be more realistic.

In the case the lower-level attack-defense process is observed (Section 4), we present new analytical results for deriving $q(v)$ under weaker assumptions, where $q(v)$ is the probability that an arbitrary node $v \in \mathsf{V}$ is compromised in the steady state. On one hand, the new analytical results lead to useful methods for obtaining the global security measure $q$ in practice. (i) In the case that $\mathsf{G}$ is regular graph with moderate node degree and the dependence between the nodes can be ignored, we derive closed-form solution for $q$. (ii) In the case that G is arbitrary graph (including the case of regular graph but with large node degree) and the dependence between the nodes can be ignored, we give an approximation solution for $q$. (iii) In the case that G is arbitrary graph and the dependence between the nodes cannot be ignored, we describe statistical solution for $q$. On the other hand, the new analytical results lead to *bounds* of $q$ in both the case G is regular graph and the case G is arbitrary graph. Such results are interesting and useful because they demand much less information. The upper bounds are especially useful in decision-making because they capture a sort of worst-case scenario from the defender's perspective.

In the case the lower-level attack-defense process is not observed, but the higher-level Alternating Renewal process is observed (Section 5), we present new analytical results for estimating $q(v)$. These analytical results lead to statistical methods for obtaining the global security measure $q$ in practice. Note that this case was not investigated in [19], and is harder to deal with analytically due to the lack of sufficient lower-level information.

Note that the results in [19] and in the present paper are applicable to both undirected and directed graphs. To simplify notations, we will focus on the case of undirected graphs, while noting that when we adapt the results to the case of directed graphs, we need to replace "node degree" with "node in-degree" (a node $v$'s in-degree is the number of nodes that pointing to $v$).

**Paper outline.** The rest of the paper is organized as follows. In Sections 2 and 3, we briefly review the model and results in [19] and some probabilistic notions, respectively. In Section 4, we present new results for the case the lower-level attack-defense process is observed. In Section 5, we present new results for the case the higher-level Alternating Renewal process is observed. In Section 6, we review related previous studies. In Section 7, we discuss the utilities of the stochastic process approach and future research directions towards the ultimate goal. In Section 8, we conclude the paper.

**Main notations.** The main notations used throughout the paper are summarized as follows.

| | |
|---:|:---|
| $\mathsf{G} = (\mathsf{V}, \mathsf{E})$ | graph abstracting a networked system with $n = |\mathsf{V}|$ nodes of average node degree $\mu$ |
| $\deg(v), q(v), K(v)$ | in the steady state, node $v \in \mathsf{V}$ is compromised with probability $q(v)$, and among $v$'s $\deg(v)$ neighbors, $K(v)$ neighbors are compromised |
| $q, K$ | in the steady state, the probability that a uniformly-chosen node is compromised is denoted by $q$, and the number of a uniformly-chosen node's compromised neighbors is denoted by $K$, where $q = \sum_{v \in \mathsf{V}} q(v)/n$ and $K = \sum_{v \in \mathsf{V}} K(v)/n$ |
| $\{(S_j(v), C_j(v)); j \geq 1\}$ | the sequence of node $v$'s cycles corresponding to the Alternating Renewal process, where each cycle consists of the time interval $S_j(v)$ during which $v$ is secure, and the time interval $C_j(v)$ during which $v$ is compromised |

## 2 A Brief Review of the Model and Results in [19]

We start with a brief review of [19]. As mentioned above, a networked system is represented as a complex network $\mathsf{G} = (\mathsf{V}, \mathsf{E})$, where $\mathsf{V}$ is the set of $n$ nodes or vertices, namely $|\mathsf{V}| = n$, that abstract (for example) computers, and $\mathsf{E}$ is the set of edges that abstract the inter-node communications/interactions that can be exploited to carry out direct attacks. At any point in time, a node $v \in \mathsf{V}$ is either secure or compromised. These abstractions were not new. The novelty of [19] is the attack-defense process over $\mathsf{G}$, which is specified by the following random variables:

- $X_1$: The time a secure node becomes compromised directly by the attacker outside of $\mathsf{G}$. This models "drive-by download"-like attacks [27].

- $X_{2,i}$: The time a secure node becomes compromised because of its $i$th compromised neighbor, $1 \leq i \leq \deg(v)$. This models attacks such as the standard malware spreading within a network.

- $Y_1$: The time a compromised node becomes secure again because the compromise is detected/cured. This models reactive defense.

- $Y_2$: The time a compromised node becomes secure again because of any other reason (e.g., patching or re-installing the software system even if the compromise is not detected). This models proactive defense.

As mentioned above, the main security measure of interest is $q$, the probability that a uniformly-chosen node is compromised in the steady state. This immediately leads to the more intuitive security measure $q \cdot n$, namely the expected number of compromised nodes in the steady state. Since in the steady sate an arbitrary node $v \in \mathsf{V}$ is compromised with probability $q(v)$, we have

$$q = \frac{\sum_{v \in \mathsf{V}} q(v)}{n}.$$

The above is perhaps the most important conceptual contribution of [19], which is very simple but insightful because $q$ captures the global property of $\mathsf{G} = (\mathsf{V}, \mathsf{E})$ in the presence of attacks and defenses. The main research task is to derive expression for $q$, which turns out to be a difficult problem because the states of the nodes are random variables that are dependent upon each other according to the complex structure of $\mathsf{G}$. As a first step, [19] focused on the special case $G$ is regular graph, namely all nodes have the same degree and the $q(v)$'s would be the same (i.e. independent of $v$). Based on the following assumption, two main results were offered in [19].

**Assumption 1** ([19])

1. Assumption on the distributions of the random variables:
   - $X_1$ follows the exponential distribution with rate $\alpha$.
   - The $X_{2,i}$'s follow the exponential distribution with rate $\gamma$.
   - $Y_1$ follows the exponential distribution with rate $\beta$.
   - $Y_2$ follows the exponential distribution with rate $\eta$.

2. Assumption on the independence between the random variables: $X_1$, $Y_1$, $Y_2$, and the $X_{2,i}$'s are independent of each other.

**Theorem 1** ([19]) *Suppose $\mathsf{G}$ is regular graph. Under Assumption 1, it holds that*

$$q = \frac{1}{1+m}, \quad \text{where} \quad m = \mathrm{E}\left[\frac{\beta + \eta}{\alpha + \gamma K}\right]$$

*and random variable $K$ indicates the number of the compromised neighbors of a node.*

**Theorem 2** ([19]) *For regular graph $\mathsf{G}$ of node degree $\mu$. Under Assumption 1, it holds that*

$$\frac{\alpha}{\alpha + \beta + \eta} \leq q \leq \frac{\alpha + \gamma\mu}{\alpha + \beta + \eta + \gamma\mu}.$$

Simulation study in [19] showed that the upper bound is reasonably tight not only for regular graphs, but also for Erdos-Renyi random graphs and power-law graphs (in the latter two cases, $\mu$ is naturally the average node degree). This means that we can plausibly use the upper bound in decision-making when there is a lack of information.

## 3 A Brief Review of Some Probabilistic Notions Used in This Paper

**Some probabilistic distributions.** A random variable $Z$ is said to have an exponential distribution with rate $\lambda > 0$ if the distribution function is

$$F(z) = 1 - \exp\{-\lambda z\}.$$

This distribution has the memoryless property, which played an essential role in [19]. Exponential distribution is a special case of the Weibull distribution with distribution function

$$F(z) = 1 - \exp\{-(\gamma z)^\alpha\},$$

where $\gamma > 0$ and $\alpha > 0$ are scale and shape parameters, respectively.

The Lomax (or Pareto II) distribution has distribution function

$$F(z) = 1 - \left(1 + \frac{z}{\beta}\right)^{-\eta},$$

where $\beta > 0$ is the scale parameter, and $\eta > 0$ is the shape parameter [14]. This distribution is interesting because it can model heavy-tailed phenomena.

**Some notions of non-independence between random variables.** Two random variables $Z_1$ and $Z_2$ are said to be Positively Quadrant Dependent (PQD) if for all $z_1, z_2 \in \mathbb{R}$,

$$P(Z_1 > z_1, Z_2 > z_2) \geq P(Z_1 > z_1)P(Z_2 > z_2).$$

In general, a family $\{Z_1, \ldots, Z_n\}$ of random variables is said to be Positively Upper Orthant Dependent (PUOD) if for all $z_i \in \mathbb{R}$, $i = 1, \ldots, n$,

$$P(Z_1 > z_1, \ldots, Z_n > z_n) \geq \prod_{i=1}^{n} P(Z_i > z_i).$$

A family $\{Z_1, \ldots, Z_n\}$ of random variables is said to be Positively Associated (PA) if

$$\text{Cov}\left(f(Z_1, \ldots, Z_n), g(Z_1, \ldots, Z_n)\right) \geq 0$$

for any coordinate-wise nondecreasing functions $f$, $g$ on $\mathbb{R}^n$ as long as the covariance exists. Finally, we recall the multivariate Marshall-Olkin model, which offers a good deal of flexibility in modeling dependence structures [21, 18]. Let $\{E_B, B \subseteq \{1, 2, \ldots, b\}\}$ be a sequence of independent, exponentially distributed random variables, where $E_B$ has mean $1/\lambda_B$. Let

$$T_j = \min\{E_B : j \in B\}, j = 1, \ldots, b.$$

The joint distribution of $T = (T_1, \ldots, T_b)$ is called the Marshall-Olkin exponential distribution with parameters $\{\lambda_B, B \subseteq \{1, 2, \ldots, b\}\}$.

It is known (see Theorem 3.2, Chapter 2 of [3]) that PA $\Longrightarrow$ PUOD(PQD), meaning that PA is (not necessarily strictly) stronger than PUOD(PQD). It is also known that Marshall-Olkin exponential distribution is PA, and hence PUOD [23].

## 4 New Results for the Case the Attack-Defense Process Is Observed

In this section, we present new analytical results for deriving $q(v)$, methods for deriving $q$ in practice, and bounds of $q$ in the absence of sufficient information.

### 4.1 New Analytical Results for Deriving $q(v)$ under Weaker Assumptions

We seek to substantially weaken the above Assumption 1, while considering more general graph structures (i.e., not just regular graphs). We start with the following assumption:

**Assumption 2** (a substantially weaker assumption)

1. Rather than assuming $X_1$, the $X_{2,i}$'s, $Y_1$ and $Y_2$ are exponential random variables, we assume:

    - $X_1$ follows an arbitrary continuous distribution $F_1$ with finite mean.
    - $X_{2,i}$ follows an arbitrary continuous distribution $F_{2,i}$ with finite mean, where the distributions may be different for different $i$'s.
    - $Y_1$ follows an arbitrary continuous distribution $G_1$ with finite mean.

- $Y_2$ *follows an arbitrary continuous distribution $G_2$ with finite mean.*

2. *Rather than assuming $X_1$, the $X_{2,i}$'s, $Y_1$, and $Y_2$ are independent of each other, we assume that $X_1$, the $X_{2,i}$'s, $Y_1$ and $Y_2$ can be dependent upon each other.*

In practice, Assumption 2 can be tested using some nonparametric methods (cf. Chapter 8 of [10]).

**Theorem 3** *Denote the joint survival function for $\{X_1, X_{2,1}, \ldots, X_{2,n}\}$ by*

$$\bar{H}_{n+1}(x_1,\ldots,x_{n+1}) = P\left(X_1 > x_1, X_{2,1} > x_2, \ldots, X_{2,n} > x_{n+1}\right),$$

*and the joint survival function for $Y_1$ and $Y_2$ by*

$$\bar{G}(y_1, y_2) = P\left(Y_1 > y_1, Y_2 > y_2\right).$$

*Under Assumption 2, the probability $q(v)$ that an arbitrary node $v \in \mathsf{V}$ is compromised in the steady state is*

$$q(v) = \frac{1}{1+m}, \tag{1}$$

*where*

$$m = \mathrm{E}\left[\int_0^\infty \bar{H}_{K(v)+1}(x,\ldots,x)dx\right] / \int_0^\infty \bar{G}(x,x)dx,$$

*and $K(v)$ is the number of the* compromised *neighbors of node $v$.*

**Proof** Observe that, in general, the states of node $v \in \mathsf{V}$ can be modeled as an Alternating Renewal process, with each cycle composed of two states: secure with random time $S_j(v)$ and compromised with random time $C_j(v)$ for $j = 1, 2, \ldots$. Under Assumption 2, for each cycle $j$ we can express $(S_j(v), C_j(v))$ as

$$S_j(v) = S(v) = \min\{X_1, \{X_{2,i}; 1 \le i \le K(v)\}\}, \quad C_j(v) = C(v) = \min\{Y_1, Y_2\},$$

where $K(v)$ is the number of $v$'s compromised neighbors. Hence,

$$\begin{aligned}
\mathrm{E}[S(v)] &= \mathrm{E}[\min\{X_1, X_{2,1}, \ldots, X_{2,K(v)}\}] \\
&= \mathrm{E}\left[\mathrm{E}\left[\min\{X_1, X_{2,1}, \ldots, X_{2,K(v)}\} \mid K(v)\right]\right] \\
&= \mathrm{E}\left[\int_0^\infty \bar{H}_{K(v)+1}(x,\ldots,x)dx\right].
\end{aligned}$$

Similarly, the time that a node is compromised in a cycle is

$$\mathrm{E}[C(v)] = \mathrm{E}[\min\{Y_1, Y_2\}] = \int_0^\infty \bar{G}(x,x)dx.$$

Note also that Assumption 2 implies that

$$\mathrm{E}[S(v) + C(v)] < \infty.$$

This and the fact that the distribution of $S(v) + C(v)$ is nonlattice guarantee (cf. Theorem 3.4.4 in [29]) that the process is steady and moreover

$$q(v) = \frac{E[C(v)]}{E[S(v)] + E[C(v)]} = \frac{\int_0^\infty \bar{G}(x,x)dx}{\mathrm{E}\left[\int_0^\infty \bar{H}_{K(v)+1}(x,\ldots,x)dx\right] + \int_0^\infty \bar{G}(x,x)dx}.$$

Hence, the desired result follows. ∎

The following corollary explains why Theorem 1 is a special case of Theorem 3.

**Corollary 1** *Theorem 1 is a special case of Theorem 3.*

**Proof** By replacing Assumption 2 with Assumption 1 and letting G be regular graph in Theorem 3, we obtain:

$$\begin{aligned} m &= \mathrm{E}\left[\int_0^\infty \bar{H}_{K(v)+1}(x,\ldots,x)dxdx\right] / \int_0^\infty \bar{G}(x,x)dx \\ &= \mathrm{E}\left[\int_0^\infty \bar{F}_1(x)\prod_{i=1}^{K(v)}\bar{F}_{2,i}(x)dx\right] / \int_0^\infty \bar{G}_1(x)\bar{G}_2(x)dx \\ &= \mathrm{E}\left[\int_0^\infty \exp\{-\alpha x\}\exp\{-K(v)\cdot\gamma\}dx\right] / \int_0^\infty \exp\{-\beta x\}\exp\{-\eta x\}dx \\ &= \mathrm{E}\left[\frac{\beta+\eta}{\alpha+\gamma K(v)}\right]. \end{aligned}$$

Therefore, the resulting $q$, namely $q(v)$ in the case of regular graphs, is the same as in Theorem 1. ∎

Now we consider the case that the random variables are independent of each other but follow the Weibull distribution (rather than the exponential distribution as required in [19]).

**Assumption 3** (weaker than Assumption 1 but stronger than Assumption 2)

1. Assumption on the distributions of the random variables:

    - $X_1$ follows the Weibull distribution with survival function $\bar{F}_1(t) = e^{-(\lambda_1 t)^\alpha}$.
    - The $X_{2,i}$'s follow the Weibull distribution with with survival function $\bar{F}_2(t) = e^{-(\lambda_2 t)^\alpha}$.
    - $Y_1$ follows the Weibull distribution with survival function $\bar{G}_1(t) = e^{-(\gamma_1 t)^\beta}$.
    - $Y_2$ follows the Weibull distribution with survival function $\bar{G}_2(t) = e^{-(\gamma_2 t)^\beta}$.

2. $X_1$, the $X_{2,i}$'s, $Y_1$, and $Y_2$ are independent of each other (as in Assumption 1).

In practice, the Weibull distribution in Assumption 3 can be tested using standard statistical techniques such as, QQ-plot, Kolmogorov-Smirnov test (cf. Chapter 14 of [17]). The following result is more general than Theorem 1 but less general than Theorem 3.

**Theorem 4** *Replacing Assumption 2 in Theorem 3 with Assumption 3, we have*

$$m = \frac{\Gamma\left(1+\frac{1}{\alpha}\right)}{\Gamma\left(1+\frac{1}{\beta}\right)}\mathrm{E}\left[\frac{\left(\gamma_1^\beta+\gamma_2^\beta\right)^{1/\beta}}{(\lambda_1^\alpha+K(v)\cdot\lambda_2^\alpha)^{1/\alpha}}\right].$$

**Proof** Note that

$$\begin{aligned} \int_0^\infty \bar{G}_1(x)\bar{G}_2(x)dx &= \int_0^\infty \exp\left\{-\left[(\gamma_1 x)^\beta+(\gamma_2 x)^\beta\right]\right\}dx \\ &= \int_0^\infty \exp\left\{-\left[\gamma_1^\beta+\gamma_2^\beta\right]x^\beta\right\}dx \\ &= \frac{1}{\left(\gamma_1^\beta+\gamma_2^\beta\right)^{1/\beta}}\Gamma\left(1+\frac{1}{\beta}\right), \end{aligned}$$

where $\Gamma(\cdot)$ is a gamma function. Note also that

$$\begin{aligned}
E\left[\int_0^\infty \bar{F}_1(x)\bar{F}_2^{K(v)}(x)dx\right] &= E\left[\int_0^\infty \exp\left\{-[(\lambda_1 x)^\alpha + K(v)\cdot(\lambda_2 x)^\alpha]\right\}dx\right] \\
&= E\left[\int_0^\infty \exp\left\{-[\lambda_1^\alpha + K(v)\cdot\lambda_2^\alpha]x^\alpha\right\}dx\right] \\
&= \Gamma\left(1+\frac{1}{\alpha}\right)E\left[\frac{1}{\lambda_1^\alpha + K(v)\cdot\lambda_2^\alpha}\right]^{1/\alpha}.
\end{aligned}$$

Using Theorem 3, we have the desired result. ∎

## 4.2 Applying the Analytical Results to Obtain $q$ in Practice

Now we show how to use the above analytical results to derive methods for obtaining $q$. While allowing $X_1$, the $X_{2,i}$'s, $Y_1$, and $Y_2$ to be possibly dependent upon each other, we further consider the (in)dependence between the states of the neighbors/nodes. This leads to three cases. (i) In the case that G is regular graph of node degree $\mu$ and the dependence between the neighbors can be ignored, we offer closed-form results in Section 4.2.1. (ii) In the case that G is arbitrary graph and the dependence between the neighbors can be ignored, we offer approximation results in Section 4.2.2. (iii) In the case that G is arbitrary graph and the dependence between the neighbors cannot be ignored, we offer statistical results in Section 4.2.3.

### 4.2.1 Regular Graphs with Ignorable Dependence Between the Nodes

In this case, any node has $\mu$ neighbors, denoted by $1,2,\ldots,\mu$, which are compromised in the steady state with probability $q_1 = q_2 = \ldots = q_\mu = q$. Let $I_i$ be a bernoulli random variable with parameter $q$, which is the probability that the $i$th neighbor is compromised. Then, $K(v) = \sum_{i=1}^\mu I_i$ is a binomial distribution with parameter $q$, and

$$P(K(v) = k) = \binom{\mu}{k}q^k(1-q)^{\mu-k}, \quad k = 0,\ldots,\mu.$$

According to Theorem 3, which requires Assumption 2, we have

$$\begin{aligned}
m &= E\left[\int_0^\infty \bar{H}_{K(v)+1}(x,\ldots,x)dx\right] / \int_0^\infty \bar{G}(x,x)dx \\
&= \sum_{k=0}^\mu \left[\binom{\mu}{k}q^k(1-q)^{\mu-k}\int_0^\infty \bar{H}_{k+1}(x,\ldots,x)dx\right] / \int_0^\infty \bar{G}(x,x)dx.
\end{aligned}$$

Hence, we have:

$$q + \sum_{k=0}^\mu \left[\binom{\mu}{k}q^{k+1}(1-q)^{\mu-k}\int_0^\infty \bar{H}_{k+1}(x,\ldots,x)dx\right] / \int_0^\infty \bar{G}(x,x)dx = 1. \quad (2)$$

In order to solve Eq. (2), we need to know the specific $\bar{H}_{k+1}(x,\ldots,x)$ and $\bar{G}(x,x)$, which can be obtained in scenarios such as the following.

- Closed-form result under Assumption 1, which is stronger than Assumption 2: In this special case, we have Theorem 1 and thus Eq. (2) becomes

$$q + \sum_{k=0}^\mu \frac{\beta+\eta}{\alpha+\gamma k}\binom{\mu}{k}q^{k+1}(1-q)^{\mu-k} = 1. \quad (3)$$

- Closed-form result under Assumption 3, which is stronger than Assumption 2: In this special case, we have Theorem 4 and thus Eq. (2) becomes

$$q + \sum_{k=0}^{\mu} \binom{\mu}{k} q^{k+1}(1-q)^{\mu-k} \frac{\Gamma\left(1+\frac{1}{\alpha}\right)}{\Gamma\left(1+\frac{1}{\beta}\right)} \left[\frac{\left(\gamma_1^\beta + \gamma_2^\beta\right)^{1/\beta}}{(\lambda_1^\alpha + k\lambda_2^\alpha)^{1/\alpha}}\right] = 1. \quad (4)$$

When $\mu$ is not large, the solutions for $q$ in Eqs. (3) and (4) can be derived using an appropriate iterative algorithm, such as the Newton-Raphson method [33]. When $\mu$ is large, it might be infeasible to derive explicit solutions for $q$, but we can instead use the approximation result in Section 4.2.2.

### 4.2.2 Arbitrary Graphs with Ignorable Dependence Between the Nodes

For an arbitrary node $v \in \mathsf{V}$, let $I_i$ be a bernoulli random variable with parameter $q_i$, which is the probability that the $i$th neighbor of node $v$ is compromised. Note that $K(v) = \sum_{i=1}^{\deg(v)} I_i$. Suppose the mean and variance of $K(v)$ exist and we can estimate them (e.g., based on sufficiently many observations on $v$'s neighbors). Denote by

$$\bar{k} = \mathrm{E}[K(v)], \quad \sigma^2 = \mathrm{Var}(K(v)).$$

In this case, we are unable to give closed-form solutions, and instead propose using two approximations.

a) Normal approximation: If $\sigma^2 \to \infty$ as $\deg(v) \to \infty$, by Lindeberg's Theorem (cf page 359 in [4]),

$$P(K(v) \le x) \approx \int_0^x \frac{1}{\sqrt{2\pi}\sigma} \exp\left\{-\frac{(y-\bar{k})^2}{2\sigma^2}\right\} dy.$$

According to Theorem 3, which requires Assumption 2, we have

$$\begin{aligned} m &= \mathrm{E}\left[\int_0^\infty \bar{H}_{K(v)+1}(x,\ldots,x)dx\right] / \int_0^\infty \bar{G}(x,x)dx \\ &\approx \int_0^\infty \int_0^{\deg(v)} \frac{1}{\sqrt{2\pi}\sigma} \exp\left\{-\frac{(k-\bar{k})^2}{2\sigma^2}\right\} \bar{H}_{k+1}(x,\ldots,x)dkdx / \int_0^\infty \bar{G}(x,x)dx. \end{aligned} \quad (5)$$

As a result, we can obtain more specific approximation solutions as follows.

- Approximation result under Assumption 1, which is stronger than Assumption 2: In this special case we have Theorem 1 and thus Eq. (5) becomes

$$m \approx \frac{\beta + \eta}{\sqrt{2\pi}\sigma} \int_0^{\deg(v)} \frac{1}{\alpha + \gamma x} \exp\left\{-\frac{(k-\bar{k})^2}{2\sigma^2}\right\} dk.$$

- Approximation result under Assumption 3, which is stronger than Assumption 2: In this special case, we have Theorem 4 and thus Eq. (5) becomes

$$m \approx \frac{\Gamma\left(1+\frac{1}{\alpha}\right)\left(\gamma_1^\beta + \gamma_2^\beta\right)^{1/\beta}}{\sqrt{2\pi}\sigma\Gamma\left(1+\frac{1}{\beta}\right)} \int_0^{\deg(v)} \frac{1}{(\lambda_1^\alpha + k\lambda_2^\alpha)^{1/\alpha}} \exp\left\{-\frac{(k-\bar{k})^2}{2\sigma^2}\right\} dk.$$

b) Poisson approximation: Note that $\bar{k} \ge 0$. If $\deg(v) \to \infty$ and $\max\{q_1,\ldots,q_{\deg(v)}\} \to 0$, Theorem 23.2 in [4] says that

$$P(K(v) = i) \approx \exp\{-\bar{k}\}\frac{(\bar{k})^i}{i!}, \quad i = 0, 1, \ldots, \deg(v).$$

According to Theorem 3, which requires Assumption 2, we have

$$m = \mathrm{E}\left[\int_0^\infty \bar{H}_{K(v)+1}(x,\ldots,x)dx\right] / \int_0^\infty \bar{G}(x,x)dx$$

$$\approx \sum_{i=0}^{\deg(v)} \exp\{-\bar{k}\} \frac{(\bar{k})^i}{i!} \int_0^\infty \bar{H}_{i+1}(x,\ldots,x)dx / \int_0^\infty \bar{G}(x,x)dx. \quad (6)$$

As a result, we can have more specific approximation solutions as follows.

- Approximation result under Assumption 1, which is stronger than Assumption 2: In this special case, we have Theorem 1 and thus Eq. (6) becomes

$$m \approx (\beta+\eta)\exp\{-\bar{k}\} \sum_{i=0}^{\deg(v)} \frac{1}{\alpha+\gamma i} \frac{(\bar{k})^i}{i!}.$$

- Approximation result under Assumption 3, which is stronger than Assumption 2: In this special case, we have Theorem 4 and thus Eq. (6) becomes

$$m \approx \frac{\Gamma\left(1+\frac{1}{\alpha}\right)\left(\gamma_1^\beta+\gamma_2^\beta\right)^{1/\beta}}{\Gamma\left(1+\frac{1}{\beta}\right)} \sum_{i=0}^{\deg(v)} \frac{(\bar{k})^i \exp\{-\bar{k}\}}{(\lambda_1^\alpha+i\cdot\lambda_2^\alpha)^{1/\alpha} i!}.$$

In both normal approximation and Poisson approximation, having obtained $m$ allows one to derive $q(v)$ according to Eq. (1) in Theorem 3. Note that this method is sound because both Assumption 1 and Assumption 3 are stronger than Assumption 2, which underlies Theorem 3. From a sufficiently large sample $\mathsf{V}' \subset \mathsf{V}$ with each $v \in \mathsf{V}'$ being uniformly chosen from $\mathsf{V}$, we can derive $q = \frac{\sum_{v\in \mathsf{V}'} q(v)}{|\mathsf{V}'|}$.

### 4.2.3 Arbitrary Graphs with Non-Ignorable Dependence Between the Nodes

If the dependence between neighbors cannot be ignored, the probability mass function of $K(v)$ may be estimated from data as follows. Suppose we have observations $(k_1, k_2, \ldots, k_b)$ on $K(v)$, where $b$ is the number of observations on the number of $v$'s compromised neighbors in the steady state (i.e., at $b$ different points of observation time). Then, the probability mass function of $K(v)$ can be estimated as

$$p_k = P(K(v) = k) = \frac{\sum_{i=1}^b \mathrm{I}(k_i = k)}{b}, \quad b\to\infty,$$

where $\mathrm{I}(\cdot)$ is the indicator function. Thus, according to Theorem 3, we have

$$m = \mathrm{E}\left[\int_0^\infty \bar{H}_{K(v)+1}(x,\ldots,x)dx\right] / \int_0^\infty \bar{G}(x,x)dx$$

$$= \sum_{k=0}^{\deg(v)} p_k \int_0^\infty \bar{H}_{k+1}(x,\ldots,x)dx / \int_0^\infty \bar{G}(x,x)dx. \quad (7)$$

To be specific, let us consider a concrete example under the following assumption:

**Assumption 4** (weaker than Assumption 1 but stronger than Assumption 2)

1. *Assumption on the distributions of the random variables:*

   - $X_1$ *follows the exponential distribution with parameter* $\lambda$.
   - *The* $X_{2,i}$*'s follow the multivariate Marshall-Olkin exponential model.*

- $Y_1$ and $Y_2$ follow the bivariate Marshall-Olkin exponential model.

2. The $X_{2,i}$'s can be dependent upon each other; $Y_1$ and $Y_2$ can be dependent upon each other.

Under Assumption 4, the joint survival function of $(X_{2,1}, \ldots, X_{2,k})$ is:

$$\begin{aligned}\bar{H}_k(x_1, \ldots, x_k) &= P(X_{2,1} > x_1, \ldots, X_{2,k} > x_k) \\ &= \exp\left\{-\sum_{i=1}^k \lambda_i x_i - \sum_{i<j} \lambda_{ij} \max\{x_i, x_j\} - \lambda_{12\ldots k} \max\{x_1, \ldots, x_k\}\right\}.\end{aligned}$$

Therefore,

$$\begin{aligned}\int_0^\infty \bar{H}_{k+1}(x, \ldots, x) dx &= \int_0^\infty \exp\{-\lambda t\} \exp\{-\phi(\boldsymbol{\lambda}, k)t\} dt \\ &= \frac{1}{\lambda + \phi(\boldsymbol{\lambda}, k)},\end{aligned}$$

where

$$\phi(\boldsymbol{\lambda}, k) = \sum_{i=1}^k \lambda_i + \sum_{i<j} \lambda_{ij} + \ldots + \lambda_{12\ldots k}.$$

Similarly, the joint survival function of $(Y_1, Y_2)$ is:

$$\begin{aligned}\bar{G}(y_1, y_2) &= P(Y_1 > y_1, Y_2 > y_2) \\ &= \exp\{-\gamma_1 y_1 - \gamma_2 y_2 - \gamma_{12} \max\{y_1, y_2\}\}.\end{aligned}$$

Then,

$$\int_0^\infty \bar{G}(x, x) dx = \frac{1}{\gamma_1 + \gamma_2 + \gamma_{12}}.$$

Using Eq (7), we have

$$m = (\gamma_1 + \gamma_2 + \gamma_{12}) \sum_{k=0}^{\deg(v)} \frac{p_k}{\lambda + \phi(\boldsymbol{\lambda}, k)}.$$

Having obtained $m$, one can derive $q(v)$ according to Eq. (1) in Theorem 3. Note that this method is sound because Assumption 4 is stronger than Assumption 2, which underlies Theorem 3. From a sufficiently large sample $V' \subset V$ with each $v \in V'$ being uniformly chosen from $V$, we can derive $q = \frac{\sum_{v \in V'} q(v)}{|V'|}$.

### 4.3 New Bounds of $q$

In the above we discussed how to obtain $q$ in various settings. In the case one cannot obtain $q$ (for example, due to the lack of sufficient information), one can instead use the bounds of $q$ in decision-making. This is plausible as long as the bounds are reasonably tight, and rational because it requires much less information to compute the bounds. In particular, one could use the upper bound of $q$ in decision-making because it can be seen as a sort of worst-case scenario. In what follows we will differentiate the case of regular graphs from the case of arbitrary graphs under the following three assumptions, respectively.

**Assumption 5** (assumption underlying our general bounds of $q$)

1. Assumption on the distributions of the random variables:

    - $X_1$ follows an arbitrary continuous distribution with survival function $\bar{F}_1$ and finite mean.
    - The $X_{2,i}$'s are PUOD with the same marginal continuous survival function $\bar{F}_2$ and finite mean.

- $Y_1$ follows an arbitrary distribution with continuous survival function $\bar{G}_1$ and finite mean.
- $Y_2$ follows an arbitrary distribution with continuous survival function $\bar{G}_2$ and finite mean.

2. $X_1$ and the $X_{2,i}$'s are independent, and $Y_1$ and $Y_2$ may be dependent upon each other.

**Assumption 6** (assumption underlying one specific bounds of $q$)

1. Assumption on the distributions of the random variables:

    - $X_1$ follows the exponential distribution with rate $\alpha$.
    - The $X_{2,i}$'s are PUOD with the same marginal distribution function $F_2$ of the exponential distribution with rate $\gamma$.
    - $Y_1$ follows the exponential distribution with rate $\beta$.
    - $Y_2$ follows the exponential distribution with rate $\eta$.

2. $X_1$ and the $X_{2,i}$'s are independent, and $Y_1$ and $Y_2$ are independent of each other.

**Assumption 7** (assumption underlying another specific bounds of $q$)

1. Assumption on the distributions of the random variables:

    - $X_1$ follows the Lomax distribution with scale parameter $\lambda$ and shape parameter $\alpha_1 > 1$;
    - The $X_{2,i}$'s are PUOD with the same marginal distribution function $F_2$ of the Lomax distribution with scale parameter $\lambda$ and shape parameter $\alpha_2 > 1$;
    - $Y_1$ follows the Lomax distribution with scale parameter $\gamma$ and shape parameter $\beta_1 > 1$;
    - $Y_2$ follows the Lomax distribution with scale parameter $\gamma$ and shape parameter $\beta_2 > 1$.

2. $X_1$ and the $X_{2,i}$'s are independent, and $Y_1$ and $Y_2$ are independent of each other.

In practice, the PUOD dependence assumption can be tested by the method suggested in [31], the exponential distribution assumption and the Lomax distribution assumption can be tested by using QQ-plot, or Kolmogorov-Smirnov test (see [17]).

### 4.3.1 Bounds of $q$ in the Case of Regular Graphs

**Theorem 5** *Suppose* G *is regular graph of node degree* $\mu$. *Under Assumption 5, in the steady state we have*

$$\int_0^\infty \bar{G}(x,x)dx \bigg/ \int_0^\infty \left[ \bar{F}_1(x) + \bar{G}(x,x) \right] dx \leq q \leq \int_0^\infty \bar{G}(x,x)dx \bigg/ \int_0^\infty \left[ \bar{F}_1(x) \bar{F}_2^{\bar{k}}(x) + \bar{G}(x,x) \right] dx,$$

*where $\bar{k}$ is the expected number of* compromised *neighbors for an arbitrary node.*

**Proof** Denote by $1, 2, \ldots, \mu$ the neighbors of an arbitrary node $v \in \mathsf{V}$. Let $I_i$ be the indicator function that $I_i = 1$ if and only if the $i$th neighbor of $v$ is compromised, and $q_i$ be the probability the $i$th neighbor is compromised. Under Assumption 5, the system will enter the steady state (see Theorem 3.4.4 in [29]). Since G is regular graph, we have $q_1 = \ldots = q_\mu = q$. In the steady state, we have

$$\bar{k} = \mathrm{E}[K(v)] = \mathrm{E}\left[\sum_{i=1}^\mu I_i\right] = q\mu.$$

Assume the $X_{2,i}$'s are PUOD with the same marginal survival function $\bar{F}_2$, then

$$
\begin{aligned}
m &= \mathrm{E}\left[\int_0^\infty \bar{F}_1(x)\bar{H}_{K(v)}(x,\ldots,x)dx\right] / \int_0^\infty \bar{G}(x,x)dx \\
&\geq \mathrm{E}\left[\int_0^\infty \bar{F}_1(x)\bar{F}_2^{K(v)}(x)dx\right] / \int_0^\infty \bar{G}(x,x)dx \\
&\geq \left[\int_0^\infty \bar{F}_1(x)\bar{F}_2^{\mathrm{E}[K(v)]}(x)dx\right] / \int_0^\infty \bar{G}(x,x)dx \\
&= \left[\int_0^\infty \bar{F}_1(x)\bar{F}_2^{\bar{k}}(x)dx\right] / \int_0^\infty \bar{G}(x,x)dx,
\end{aligned}
\qquad (8)
$$

where the last inequality follows from Jensen's inequality. Using Eq. (1) in Theorem 3, one can obtain the upper bound of $q$.

On the other hand, we observe that, for any $x \geq 0$,

$$\bar{H}_{K(v)}(x,\ldots,x)dx \leq 1.$$

The upper bound of $m$ can be derived as

$$m \leq \int_0^\infty \bar{F}_1(x)dx / \int_0^\infty \bar{G}(x,x)dx. \qquad (9)$$

Using Eq. (1) in Theorem 3, one can obtain the lower bound of $q$. ∎

**Theorem 6** *Suppose* G *is regular graph of node degree* $\mu$. *Under Assumption 6, in the steady state we have*

$$\frac{\alpha}{\alpha+\beta+\gamma} \leq q \leq \frac{-(\beta+\eta+\alpha-\mu\gamma)+\sqrt{(\beta+\eta+\alpha-\mu\gamma)^2+4\mu\gamma\alpha}}{2\mu\gamma}.$$

**Proof** From Eq. (8), it follows that,

$$m \geq \frac{\beta+\eta}{\alpha+q\mu\gamma}.$$

Using Eq. (1) in Theorem 3, we have

$$q = \frac{1}{m+1} \leq \frac{1}{\frac{\beta+\eta}{\alpha+q\mu\gamma}+1}.$$

This leads to

$$q^2\mu\gamma + q(\beta+\eta+\alpha-\mu\gamma) - \alpha \leq 0.$$

Hence,

$$q \leq \frac{-(\beta+\eta+\alpha-\mu\gamma)+\sqrt{(\beta+\eta+\alpha-\mu\gamma)^2+4\mu\gamma\alpha}}{2\mu\gamma}.$$

On the other hand, according to Eq. (9) and Eq. (1), one can derive

$$q \geq \frac{\alpha}{\alpha+\beta+\gamma}.$$

∎

**Theorem 7** *Suppose* G *is regular graph of node degree* $\mu$. *Under Assumption 7, in the steady state we have*

$$\frac{\gamma(\alpha_1-1)}{\lambda(\beta_1+\beta_2-1)+\gamma(\alpha_1-1)} \leq q \leq \frac{-B+\sqrt{B^2+4\gamma^2\alpha_2(\alpha_1-1)\mu}}{2\alpha_2\gamma\mu},$$

*where* $B = \lambda(\beta_1+\beta_2-1)+\gamma(\alpha_1-1)-\gamma\alpha_2\mu$.

**Proof** Note that $F_i$ has Lomax survival function

$$\bar{F}_i(x) = \left(1 + \frac{x}{\lambda}\right)^{-\alpha_i}, \alpha_i \geq 1, \quad i = 1, 2.$$

Similarly, $G_i$ has Lomax survival function

$$\bar{G}_i(x) = \left(1 + \frac{x}{\gamma}\right)^{-\beta_i}, \quad \beta_i \geq 1, \quad i = 1, 2.$$

From Eq. (8), it follows that,

$$m \geq \frac{\lambda}{\alpha_1 + \alpha_2 \mu q - 1} \Big/ \frac{\gamma}{\beta_1 + \beta_2 - 1}$$
$$= \frac{\lambda(\beta_1 + \beta_2 - 1)}{\gamma(\alpha_1 + \alpha_2 \mu q - 1)}.$$

Using Eq. (1) in Theorem 3, we have

$$q \leq \frac{\gamma(\alpha_1 + \alpha_2 \mu q - 1)}{\lambda(\beta_1 + \beta_2 - 1) + \gamma(\alpha_1 + \alpha_2 \mu q - 1)}.$$

This leads to

$$q^2 \alpha_2 \gamma \mu + qB - \gamma(\alpha_1 - 1) \leq 0,$$

where

$$B = \lambda(\beta_1 + \beta_2 - 1) + \gamma(\alpha_1 - 1) - \gamma \alpha_2 \mu.$$

Hence, the upper bound of $q$ is

$$q \leq \frac{-B + \sqrt{B^2 + 4\gamma^2 \alpha_2 (\alpha_1 - 1)\mu}}{2\alpha_2 \gamma \mu}.$$

On the other hand, according to Eq. (9) and Eq. (1), we have

$$m \leq \frac{\lambda(\beta_1 + \beta_2 - 1)}{\gamma(\alpha_1 - 1)}.$$

So, the lower bound of $q$ is

$$q \geq \frac{\gamma(\alpha_1 - 1)}{\lambda(\beta_1 + \beta_2 - 1) + \gamma(\alpha_1 - 1)}.$$

■

### 4.3.2 Bounds of $q$ in the Case of Arbitrary Graphs

For an arbitrary node $v \in V$, it follows that

$$K(v) = \sum_{i=1}^{\deg(v)} I_i,$$

where $I_i$ is a bernoulli random variable with parameter $q_i$, which is the probability that the $i$th neighbor of $v$ is compromised. Note that the $I_i$'s may be dependent upon each other and the $q_i$'s may be different. Note also that in this case we cannot obtain a counterpart of Theorem 5 because the concavity needed to apply Jensen's inequality may not hold. As such, we are only able to give the following specific results in this more general case.

**Theorem 8** *Suppose* G *is arbitrary graph of with average node degree $\mu$. Under Assumption 6, in the steady state we have*

$$\frac{\alpha}{\alpha + \beta + \gamma} \leq q \leq \frac{\alpha + \gamma \mu}{\alpha + \beta + \eta + \gamma \mu}.$$

**Proof** In a fashion similar to the proof of Theorem 5, we can show that for an arbitrary node $v$,

$$\frac{\alpha}{\alpha + \beta + \gamma} \leq q(v) \leq \frac{\alpha + \gamma \mathrm{E}[K(v)]}{\alpha + \beta + \eta + \gamma \mathrm{E}[K(v)]}.$$

Since

$$q = \frac{\sum_{v \in \mathsf{V}} q(v)}{n},$$

it follows that

$$q \geq \frac{\alpha}{\alpha + \beta + \gamma}.$$

On the other hand, because

$$\mathrm{E}[K(v)] = \mathrm{E}\left[\sum_{i=1}^{\deg(v)} I_i\right] = \sum_{i=1}^{\deg(v)} q_i \leq \deg(v),$$

we have

$$\frac{\alpha + \gamma \mathrm{E}[K(v)]}{\alpha + \beta + \eta + \gamma \mathrm{E}[K(v)]} \leq \frac{\alpha + \gamma \deg(v)}{\alpha + \beta + \eta + \gamma \deg(v)}.$$

Hence,

$$q(v) \leq \frac{\alpha + \gamma \deg(v)}{\alpha + \beta + \eta + \gamma \deg(v)}.$$

So, it holds that

$$q \leq \frac{1}{n} \sum_{v \in \mathsf{V}} \frac{\alpha + \gamma \deg(v)}{\alpha + \beta + \eta + \gamma \deg(v)}. \tag{10}$$

Because

$$h(x) = \frac{\alpha + \gamma x}{\alpha + \beta + \eta + \gamma x}, \quad x \geq 0,$$

is a concave function, Jensen's inequality says

$$\frac{1}{n} \sum_{v \in \mathsf{V}} \frac{\alpha + \gamma \deg(v)}{\alpha + \beta + \eta + \gamma \deg(v)} \leq \frac{\alpha + \gamma \frac{1}{n} \sum_{v \in \mathsf{V}} \deg(v)}{\alpha + \beta + \eta + \gamma \frac{1}{n} \sum_{v \in \mathsf{V}} \deg(v)}.$$

Hence, by Eq. (10), we have

$$q \leq \frac{\alpha + \gamma \mu}{\alpha + \beta + \eta + \gamma \mu}.$$

This completes the proof. ∎

In a similar fashion, we can prove the following result.

**Theorem 9** *Suppose* G *is an arbitrary graph of with average degree* $\mu$. *Under Assumption 7, in the steady state we have*

$$\frac{\gamma(\alpha_1 - 1)}{\lambda(\beta_1 + \beta_2 - 1) + \gamma(\alpha_1 - 1)} \leq q \leq \frac{\gamma(\alpha_1 + \alpha_2 \mu - 1)}{\lambda(\beta_1 + \beta_2 - 1) + \gamma(\alpha_1 + \alpha_2 \mu - 1)}.$$

## 5 New Results for the Case Only the Alternating Renewal Process Is Observed

The above results are applicable when the lower-level attack-defense process is observed, or more specifically the distributions $F_1$, the $F_{2,i}$'s, $G_1$ and $G_2$ and their dependence structure (if applicable) are observed. What if they are not observed? This is a very reasonable question because obtaining all the detailed information can be very costly, and it may be much simpler and easier to observe the higher-level Alternating Renewal process $\{(S_j(v), C_j(v)); j \geq 1\}$ mentioned above. In this case, neither the results in [19] nor the results described above are applicable, which motivates us to present the following results for this possibly more realistic scenario.

## 5.1 Analytical Results for Estimating $q(v)$

Let $p_v(t)$ be the probability that node $v$ is secure at time $t$, and the limiting average probability be

$$\bar{p}(v) = \lim_{t \to \infty} \frac{1}{t} \int_0^t p_v(s) ds.$$

For node $v \in \mathsf{V}$ with $\{(S_j(v), C_j(v)) : 1 \leq j \leq b\}$, define

$$W_b(v) = S_1(v) + S_2(v) + \ldots + S_b(v), \quad D_b(v) = C_1(v) + C_2(v) + \ldots + C_b(v).$$

**Lemma 1** (Theorem 1 in [20]) If

$$\lim_{b \to \infty} \frac{W_b(v)}{b} = w(v) \quad a.s. \quad \text{and} \quad \lim_{b \to \infty} \frac{D_b(v)}{b} = d(v) \quad a.s.,$$

where "a.s." means "almost surely", and $0 < w(v), d(v) < \infty$. Then,

$$\bar{p}(v) = \frac{w(v)}{w(v) + d(v)}.$$

**Assumption 8** *Assume $\{S_j(v) : j \geq 1\}$ and $\{C_j(v) : j \geq 1\}$ are sequences of pairwise PQD random variables with finite variances, and*

*a)* $\sum_{j=1}^\infty j^{-2} \mathrm{Cov}\,(S_j(v), W_j(v)) < \infty$;

*b)* $\sum_{j=1}^\infty j^{-2} \mathrm{Cov}\,(C_j(v), D_j(v)) < \infty$;

*c)* $\sup_{j \geq 1} \mathrm{E}|S_j(v) - \mathrm{E}[S_j(v)]| < \infty$;

*d)* $\sup_{j \geq 1} \mathrm{E}|C_j(v) - \mathrm{E}[C_j(v)]| < \infty$.

In practice, the above assumption can be tested as follows: PQD dependence can be tested by using the statistical methods described in [7]. To test for the finite variance, we may use the methods described in [28] for heavy-tailed data. If the data does not show extremely heavy-tailed property, we may assume that the data has finite variance.

**Lemma 2** (Theorem 1 in [5]) Let $\{S_j(v) : j \geq 1\}$ be sequence of pairwise PQD random variables with finite variance. Assume

*a)* $\sum_{j=1}^\infty j^{-2} \mathrm{Cov}\,(S_j(v), W_j(v)) < \infty$.

*b)* $\sup_{j \geq 1} \mathrm{E}|S_j(v) - \mathrm{E}[S_j(v)]| < \infty$.

Then, as $b \to \infty$, we have $b^{-1}(W_b(v) - \mathrm{E}[W_b(v)]) \to 0$ almost surely.

**Theorem 10** *Under Assumption 8 and given sufficiently large $b$ observations on the states of node $v$, the probability $q(v)$, namely the probability node $v$ is compromised, if existing (i.e., the Alternating Renewal Process is steady), can be expressed as*

$$q(v) = \frac{d^*(v)}{w^*(v) + d^*(v)},$$

*where $w^*(v) = \lim_{b \to \infty} \mathrm{E}[W_b(v)]/b$ and $d^*(v) = \lim_{b \to \infty} \mathrm{E}[D_b(v)]/b$.*

**Proof** According to Lemma 2, conditions a) and c) imply

$$\lim_{b\to\infty} \frac{W_b(v)}{b} = \lim_{b\to\infty} \frac{\mathrm{E}[W_b(v)]}{b} = w^*(v), \quad a.s.,$$

and similarly, b) and d) imply

$$\lim_{b\to\infty} \frac{D_b(v)}{b} = \lim_{b\to\infty} \frac{\mathrm{E}[D_b(v)]}{b} = d^*(v), \quad a.s..$$

According to Lemma 1, it holds that

$$\bar{p}(v) = \frac{w^*(v)}{w^*(v) + d^*(v)}.$$

The limiting probability is

$$p(v) = \lim_{t\to\infty} p_v(t),$$

which means there exists a sufficiently large $t_0$ such that for any $\epsilon > 0$,

$$|p(v) - p_v(t)| < \epsilon, \quad t \geq t_0.$$

Now, for $t > t_0$, we have

$$\frac{1}{t}\int_0^t p_v(s)ds \leq \frac{1}{t}\int_0^{t_0} p_v(s)ds + \frac{1}{t}\int_{t_0}^t p_v(s)ds.$$

Hence,

$$\bar{p}(v) = \lim_{t\to\infty} \frac{1}{t}\int_0^t p_v(s)ds \leq p(v) + \epsilon.$$

Similarly, it holds that,

$$\bar{p}(v) \geq p(v) - \epsilon.$$

So, we conclude that

$$p(v) = \bar{p}(v) = \frac{w^*(v)}{w^*(v) + d^*(v)},$$

and

$$q(v) = 1 - p(v) = \frac{d^*(v)}{w^*(v) + d^*(v)}.$$

∎

In what follows, we present two variants of Theorem 10. In the first variant, we use the stronger PA assumption to replace the PQD assumption in Theorem 10 (or more precisely, in the underlying Assumption 8), which leads to that conditions c) and d) can be dropped. To see this, we need to use the following lemma.

**Lemma 3** (Theorem 2 in [5]) *Let $\{S_j(v) : j \geq 1\}$ be are sequence of pairwise PA random variables with finite variance. Assume that*

$$\sum_{j=1}^{\infty} j^{-2}\mathrm{Cov}\,(S_j(v), W_j(v)) < \infty.$$

*Then, as $b \to \infty$, we have $b^{-1}(W_b(v) - \mathrm{E}[W_b(v)]) \to 0$ almost surely.*

**Assumption 9** *Assume $\{S_j(v) : j \geq 1\}$ and $\{C_j(v) : j \geq 1\}$ are both sequences of pairwise PA random variables with finite variances, and*

a) $\sum_{j=1}^{\infty} j^{-2}\mathrm{Cov}\,(S_j(v), W_j(v)) < \infty$;

b) $\sum_{j=1}^{\infty} j^{-2}\mathrm{Cov}\,(C_j(v), D_j(v)) < \infty$;

**Theorem 11** (one variant of Theorem 10) *Under Assumption 9 and given sufficient large b observations on node v's states, the probability $q(v)$, if existing (i.e., the Alternating Renewal Process is steady), can be expressed as*

$$q(v) = \frac{d^*(v)}{w^*(v) + d^*(v)}.$$

**Proof** The proof follows from Lemma 3 and the proof of Theorem 10. ∎

In the following we present another useful variant of Theorem 10 under another stronger assumption.

**Assumption 10** *Assume $\{S_j(v) : j \geq 1\}$ and $\{C_j(v) : j \geq 1\}$ are both independent sequences with finite variances.*

In practice, we can test the independence for sequences $\{S_j : j \geq 1\}$ and $\{C_j : j \geq 1\}$ using the methods described in [10].

**Theorem 12** (another variant of Theorem 10) *Under Assumption 10 and given sufficiently large b observation on node v's states, the probability $q(v)$, if existing (i.e., the Alternating Renewal Process is steady), it can be expressed as*

$$q(v) = \frac{d^*(v)}{w^*(v) + d^*(v)}.$$

**Proof** The theorem follows from Kolmogorov's strong law of large numbers and the proof of Theorem 10. ∎

The following corollary demonstrates the connection between Theorem 12, which corresponds to the case the higher-level Alternating Renewal Process is observed, and Theorem 1, which corresponds to the case the lower-level attack-defense process is observed.

**Corollary 2** *Theorem 1 is a special case of Theorem 12.*

**Proof** Under Assumption 1 used in Theorem 1, we have

$$\mathrm{Var}(S(v)) < 2\mathrm{E}\left[\frac{1}{\alpha + K(v) \cdot \gamma}\right]^2 \leq 2\left(\frac{1}{\alpha}\right)^2 < \infty, \quad \mathrm{Var}(C(v)) = \left(\frac{1}{\beta + \eta}\right)^2 < \infty.$$

Hence, the conditions in Theorem 12 are fulfilled. Therefore, we have

$$w^*(v) = \mathrm{E}[S(v)] = \mathrm{E}\left[\frac{1}{\alpha + \gamma K(v)}\right], \quad d^*(v) = \mathrm{E}[C(v)] = \frac{1}{\beta + \eta}.$$

As a result,

$$q(v) = \mathrm{E}\left[\frac{\beta + \eta}{\alpha + \gamma K(v)}\right],$$

which is the same as in Theorem 1. ∎

## 5.2 Applying the Analytical Results to Estimate $q$ in Practice

Theorems 10-12 only require observations on $\{(S_l(v), C_l(v)); l \geq 1\}$. In order to show how the above analytical results can guide the computation of $q$ in practice, we use Theorem 12 as an example. Suppose we select $m$ nodes, denoted by $v_1, \ldots, v_m$, uniformly at random from V to observe their respective $\{(S_l(v), C_l(v)); l \geq 1\}$'s. For each $v_i$ where $1 \leq i \leq m$, we estimate $q(v_i)$ as follows:

1. Observe the node for a sufficient long time,[1] and get $j_i$ observations $(S_1(v_i), C_1(v_i)), \ldots, (S_{j_i}(v_i), C_{j_i}(v_i))$.

---
[1] In practice, we can first observe time interval $[0, t_1]$ and use Eq. (11) to compute $q(v_i)$. Then, we can observe a longer interval, say $[0, t_2]$ and compute $q(v_i)$ again. Assume we collect many points $(q(v_i), t_z), z = 1, 2, \ldots$. Then, we may plot them to observe the horizontal line. If the horizontal line pattern appears after $t_z$ for some $z$, we may say $t_z$ is sufficient long time. This method can also be used to determine whether the process is steady in practice.

2. Test whether $\{S_l(v_i) : j_i \geq l \geq 1\}$ and $\{C_l(v_i); j_i \geq l \geq 1\}$ are independent sequences (e.g., using methods described in [10]). If they are independent, the algorithm continues executing the next steps; otherwise, the algorithm aborts.

3. Test whether $\{S_l(v_i) : j \geq l \geq 1\}$ and $\{C_l(v_i); j \geq l \geq 1\}$ have finite variances (e.g., using the methods described in [28]). If so, the algorithm continues executing the next step; otherwise, the algorithm aborts.

4. Compute $W_{i,j} = S_1(v_i) + \ldots + S_j(v_i)$ and $D_{i,j} = C_1(v_i) + \ldots + C_j(v_i)$.

5. Compute
$$q(v_i) = \frac{D_{i,j}}{W_{i,j} + D_{i,j}}. \tag{11}$$

Having obtained $q(v_1), \ldots, q(v_m)$, we can compute
$$\bar{q} = \frac{\sum_{i=1}^{m} q(v_i)}{m},$$
which, while intuitive, is not unconditionally applicable as our algorithm demonstrates.

## 6 Related Work

As extensively discussed in [19], there have been some (promising) attempts at quantitative security analysis of networked systems. In what follows we discuss three main approaches, which are complementary to the *stochastic process* approach of [19] and the present paper.

The extensively studied approach is the *attack graph* approach, which was initiated by [26]. Basically, an attack graph is derived from a networked system and its known vulnerabilities. The nodes in an attack graph are the states of the networked system and the edges reflect the attack steps (i.e., having compromised certain computers can cause the compromise of more computers). Along this line, substantial progresses have been made in the last decade [13, 32, 2, 24, 12, 25, 36, 30, 34, 37, 39, 11]. A particular application of attack graphs is to (optimally) harden specific assets by identifying the relevant attack paths. Our approach is complementary to the attack graph approach because of the following. First, the attack graph approach mainly focuses on studying known (but unpatched) vulnerabilities. The only exception we are aware of is the recent extension to consider the number of unknown vulnerabilities that are needed in order to compromise some specific assets [35]. In contrast, our approach aims to embed unknown vulnerabilities into the model inputs (rather than as an outcome of the model). Second, the attack graph approach is algorithmic or combinatorial in nature (e.g., computing the number of attack paths or hardening security of a given set of assets with minimal effort). In contrast, our approach aims to model the evolution of node states, which could allow us to derive basic laws/principles for the security of networked systems.

The approach that is closely related to ours is what we call the *dynamical system* approach, in which Dynamical System models play an essential role. This approach was rooted in epidemic models [22, 16], which were first introduced to computer security for studying computer viruses in [15]. This approach has been coupled with complex network structures since [38, 8, 6]. The state-of-the-art result in this line of research is [40], which also presented perhaps the first numerical result for estimating the global state that is comparable to our concept of $q$. While the results in [40] can accommodate arbitrary network structures, they are based on an independence assumption that is comparable to that the $X_{2,i}$'s in our model are independent of each other. As shown above, we aim to get rid of this independence and other assumptions as much as we can.

The third approach is what we call the *statistical physics* approach (cf. [1] for a large body of literature and its numerous follow-ons). This approach mainly aims to characterize the robustness of network connectivity (i.e., robustness of giant components in the presence of node and/or edge deletions). While relevant, there is a fundamental difference between network connectivity and security. Specifically, as noted in [19], a malicious attacker may aim to compromise as many computers as possible. However, the ultimate goal of the attacker is to steal sensitive information, rather than to disrupt the connectivity of networks. In other words, the attacker will compromise sensitive information, while likely not disrupting the communications of the legitimate users.

# 7  Discussion and Research Directions

In this Section we discuss the utilities of the stochastic process approach, and suggest future research directions.

**On the utilities of the stochastic process approach.** The stochastic process approach was introduced by [19], which can be seen as a new way of thinking about quantitative security. The present paper is a first step towards eliminating the exponential-distribution assumptions and the independence assumptions in [19]. The approach is centered at stochastic processes that model the interaction between the attacks and the defense. The approach is at its infant stage and is currently based on first-principle modeling (due to the lack of real-life data). Nevertheless, the approach has important real-life implications as we elaborate below.

First, since the stochastic processes capture the evolution of node states, the results are pertinent to the evolution of system states. This will help deepen our understanding of the problem and likely will allow us to draw insights or laws/principles that govern the outcome of the interaction between the attacks and the defense. Note that the characterization is not fundamentally based on knowledge of the parameters because the parameters *do exist* anyway. In other words, the resulting insights or laws/principles are valid, whether we know the specific values of the model parameters or not. Such general (or even universal) laws/principles are of paramount importance.

Second, the analytical results could lead to guidelines (as shown in [19]) for adjusting the defense in a cost-effective, if not optimal, fashion. As mentioned above, the bounds can be utilized in practice to capture the worst-case scenario, which can be adopted for decision-making. For example, if we know that $q$ is bounded from above, then some appropriate proactive threshold cryptosystems [9] may be deployed to tolerate the bounded compromise. Moreover, the study can suggest quantitative defense adjustments so as to contain $q$ below a desired threshold (as shown in [19]).

Third, in the above we offered statistical methods for obtaining security measures in practice. In Section 4, we presented numerical methods for deriving the global security measure $q$ in various practical settings. In Section 5, we presented statistical methods for obtaining the global security measure $q$ when only high-level information about the attack-defense process can be obtained.

**Extending the stochastic process models.** The models in the present paper can be extended in multiple directions. The first direction is to further weaken the assumptions as much as possible. One question is: How should we accommodate node- and edge-heterogeneity (other than the degree- or topology-heterogeneity), meaning that different nodes (edges) exhibit different parameter distribution characteristics? Note that our analytical results on $q(v)$, the probability that node $v$ is compromised in the steady state, already accommodated node- and edge-heterogeneity. However, our analytical results on the global security measure $q$, especially its bounds, are based on that the nodes (edges) exhibit the same parameter distribution characteristics. Another question is: How can we bridge the gap between what can be observed in practice (i.e., what can be offered by real-life data) and the weakest possible assumptions that we have to make in order to derive analytical results? This includes the design of new statistical methods for testing assumptions that currently cannot be tested.

The second direction is to accommodate multiple vulnerability classes but from a different perspective. Our models are based on aggregating the effects of vulnerabilities at individual computers. Another perspective is to consider individual vulnerability classes that are "compatible" with each other, where *compatibility* is an intuitive concept that needs to be formalized. The intuition is that the exploitation of some vulnerability in one computer can cause the exploitation of another compatible (i.e., not necessarily the same) vulnerability in another computer. This would lead to multiple stochastic attack-defense processes that might be dependent upon each other. This quickly becomes very complicated, but certainly worthy of study because it may lead to deeper insights. Moreover, it is certainly interesting and important to compare the two perspectives.

The third direction is to accommodate that the consequence-heterogeneity between the nodes. This is important because the damage that is caused by the compromise of one node (e.g., a server) may be greater than the damage that is caused by the compromise of another (e.g., a desktop). Our models can partially accommodate this consequence-heterogeneity via the $q(v)$'s, where $q(v)$ is the probability that node $v$ is compromised in the steady state. However, our results on, including the bounds of, the global security measure $q$ do not consider the asset on node $v$, which may be denoted by $asset(v)$. One simple extension would be to introduce a derivative security measure $q(v) \times asset(v)$, which can lead to the naturally extended global security measure $\frac{\sum_{v \in V} q(v) \times asset(v)}{|V|}$. Nevertheless, other extensions are possible.

The fourth direction is to characterize the distribution of the random variable that indicates the number of compromised nodes in the steady state. Note that $q \times |V|$ corresponds to the expected number of compromised nodes in the steady state. Knowledge about the distribution of the random variable will allow us to conduct better decision-making.

**Unifying the approaches.** As mentioned in the Introduction, the problem of quantitative security analysis of networked computer systems has been outstanding for decades. In the above Related Work section, we highlighted three main approaches. Both the attack graph approach and the statistical approach have been extensively studied, but both the dynamical system approach and the stochastic process approach are at their infant stages. Nevertheless, it is imperative to unify these approaches into a comprehensive and systematic framework.

## 8 Conclusion

We extended the quantitative security model and analysis presented in [19] by substantially weakening its assumptions regarding the distributions of, and the dependence between, the relevant random variables. In particular, our extensions led to practical methods for obtaining the desired security measures for both the case that the lower-level attack-defense process is observed and the case that the higher-level Alternating Renewal process is observed. We discussed future research directions towards ultimately solving the problem of quantitative security analysis of networked systems.

**Acknowledgement.** We are very grateful to an anonymous reviewer whose comments helped us improve the paper.

Shouhuai Xu's work was supported in part by grants sponsored by AFOSR, AFOSR MURI and NSF. The views and conclusions contained in the article are those of the authors and should not be interpreted as, in any sense, the official policies or endorsements of the US government or the agencies.